\documentclass[twocolumn]{jpsj3}
\usepackage[dvips]{color}
\usepackage{txfonts}
\usepackage{cases}

\title{
Topological Quantum Phase Transitions in a Majorana Chain with Spatial Modulation
}

\author{Takumi Ohta\thanks{takumi@yukawa.kyoto-u.ac.jp} and Keisuke Totsuka}
\inst{
Yukawa Institute for Theoretical Physics, Kyoto University, Kyoto 606-8502, Japan} 

\abst{
We numerically study the quantum phase transitions and the stability of Majorana zero modes
in a generalized Kitaev model in one dimension when the chemical potential is periodically modulating in space.
By using the exact diagonalization method for open boundary condition,
we investigate the ground-state phases in terms of the non-local properties
such as the entanglement spectrum (ES) and the string correlation functions.
When we vary the phase of the modulation, the number of the Majorana zero modes changes,
which manifests itself in the degeneracy of the lowest level of the ES.
Next, we study the quantum phase transitions driven by the change in the amplitude of the modulation.
In particular, for certain values of the wave number and the phase of the modulation,
we observe a quantum phase transition from one topological phase into another
where the string correlation function oscillates in space.
We also show a case where the degeneracy of the ES does not change even for large enough amplitude of the modulation.
Finally, we characterize the phases of the system with periodic boundary condition by the topological invariant,
which reflects the number of the zero-energy excitations.
}

\begin{document}

\maketitle

\section{\label{sec:Introduction}Introduction}
Motivated by the recent development in mesoscopic physics and quantum-information science,
both theoretical and experimental studies have discovered topological phases 
such as quantum Hall states,\cite{Klitzing-1980,Laughlin-1981} 
topological insulators,\cite{Kane-2005,Konig-2007} and quantum spin liquids.\cite{Wen-1989,Wen-1991}
These topological phases are characterized not by any local order parameters
but by non-local order parameters,\cite{Nijs-1989,Hatsugai-1991,Kennedy-1992,Totsuka-1995}
their emergent edge excitations,\cite{Affleck-1987,Wen-1990}
and entanglement.\cite{Kitaev-2006,Levin-2006,Furukawa-2007,Li-2008}
The non-local order parameters\cite{Nijs-1989} for the Haldane phase in the spin 1 Heisenberg antiferromagnet
are called the string order parameters
and are able to detect the hidden antiferromagnetic order in the ground state.\cite{Affleck-1987}
The concept of entanglement also plays an important role to characterize the topological phases.
Li and Haldane have proposed the use of the entanglement spectrum (ES) to quantify entanglement
and found that the ES of fractional quantum Hall states has a close relationship with
the energy spectrum of the low-lying excitations appearing at the edge of the system.\cite{Li-2008}
Thereafter the ES has been used in theoretical studies of the topological phases
and it is known that the level structure of the ES reflects that of the edge modes of the topological phases.
\cite{Pollmann-2010,Cirac-2011,Lou-2011,Fagotti-2011,Tanaka-2012,Qi-2012,Nonne-2013,Cincio-2013,Tanimoto-2015}

There are topological phases which have Majorana fermions as the quasiparticle excitation.
One of the simplest models which are known to host the phases
would be the Kitaev model of spinless $p$-wave superconductor in one dimension.\cite{Kitaev-2001}
This model is realized by the proximity effect of a quantum wire on superconductor.\cite{Oreg-2010,Mourik-2012,Rokhinson-2012}
A similar model is obtained using the optically trapped fermion atoms.\cite{Jiang-2011}
In the topological phase,
a pair of the Majorana zero modes appearing at the ends of the system form a non-local fermion excitation with zero energy,
leading to the two-fold degeneracy in the ground states.
The modes are stable until the energy gap of the system closes and thus characterize the topological phase.
Because of the non-local nature of the Majorana fermions,
they obey non-Abelian statistics and can be used for the topological quantum computation.\cite{Bravyi-2002,Nayak-2008}

The study of the Majorana fermions is important from the view point of both quantum information science and condensed matter physics.
Since the first proposal of quantum computation using the Majorana zero modes,\cite{Bravyi-2002}
the stability of the modes has been attracting much attention from the quantum-information perspective.\cite{Alicea-2011,Wu-2014,Amorim-2015}
To manipulate them in the actual experiments, we must know entanglement and non-local correlation of them.
In condensed matter physics, on the other hand,
much effort has been devoted to the investigation of the topological phases exhibiting the Majorana fermions.\cite{Alicea-2012}
There are theoretical studies concerning the effect of
periodic modulation,\cite{Lang-2012,Degottardi-2013,Degottardi2-2013,Cai-2013,Tezuka-2013}
long-range (i.e. further-neighbor) interactions,\cite{Ohta-2015,Ohta-2016}
and quartic interaction.\cite{Stoudenmire-2011}
The stability of the Majorana zero modes against spatially periodic modulation and disorder
has been considered in Ref.~\citen{Lang-2012},
where the authors found out the enhancement of the topological phase by the modulation and
the transitions from the topological phase to a trivial phase.
In Refs.~\citen{Ohta-2015,Ohta-2016}, we considered a model with several types of Majorana interactions such as long-range ones
which may change the number of the Majorana zero modes.
Focusing on entanglement and non-local correlation,
we reveal a variety of phases (both topological and non-topological) resulting from the competition among the interactions
that affect the pattern of the correlation of the Majorana fermions.
Because the modulation of the chemical potential influences the topological phases
by inhomogeneously disturbing the pairing pattern of the Majorana fermions,
the interplay of the long-range interaction and the modulating chemical potential
would cause non-trivial effects on the formation of topological phases characterized by the Majorana fermions.

In this paper,
we investigate the quantum phase transitions in a generalized Kitaev model with spatially periodic modulation in one dimension
focusing on entanglement and non-local correlation.
Specifically, we consider the effect of a next-nearest-neighbor interaction and the periodic modulation in the chemical potential
on the topological properties of the Kitaev model.
In principle, we could include the nearest-neighbor interaction as well.
However, to avoid complexity and understand the essence of the physical picture,
we only consider the next-nearest-neighbor interaction.
To this end, we numerically calculate the ES and the string correlation functions
by using the exact diagonalization method.
When we vary the modulation, the degeneracy in the lowest level of the ES changes,
which we confirm that corresponds to the degeneracy of the ground states.

The rest of this paper is organized as follows:
In Sec.~\ref{sec:Model}, we introduce our model and give its Majorana representation.
We sketch how to map out the ground-state phases in Sec.~\ref{sec:Methods},
where we also introduce the correlation functions and the ES to characterize the topological phases.
In Sec.~\ref{sec:Quantum phase transitions}, we study the quantum phase transitions of the model with open boundary condition
driven by changing the amplitude, the wave number, and the phase of the periodic modulation.
We characterize each phase by the string correlation functions and the ES.
Focusing on the topological phases,
we discuss the topological phase transitions and the stability of the Majorana edge modes.
In Sec.~\ref{sec:Topological invariant}, we characterize each phase by the topological invariant
which is given by an integral over the momentum space.
In Sec.~\ref{sec:Summary}, we summarize our results and conclude this paper.

\section{\label{sec:Model}Model}
\begin{figure}
 \begin{center}
  \includegraphics[width=86mm]{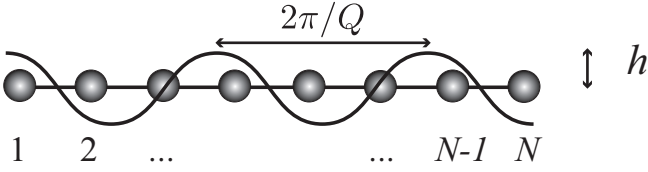}
 \end{center}
 \caption{(Color online)
 One-dimensional chain [Eq.~(\ref{eq:Hquad})] of $N$ sites with spatially periodic modulation
 whose amplitude, the wave number, and the phase are $h$, $Q$, and $\delta$ respectively.
 The black circles mean the sites where the fermions are defined.
 }
 \label{fig:Fig1}
\end{figure}

We consider a generalized Kitaev model in one dimension with spatially periodic modulation (Fig.~\ref{fig:Fig1}):
\begin{equation}
\label{eq:Hth}
H=\sum_{i=1}^N\left(t c_i^{\dagger}c_{i+2} +t c_i^{\dagger}c_{i+2}^{\dagger} +{\rm h.c.} -2h\cos(Qi+\delta) c_i^{\dagger}c_i   \right),
\end{equation}
where $N$ and $t$ respectively are the system size and the hopping amplitude.
The periodic modulation of the chemical potential is controlled
by the amplitude $h$, the wave number $Q$, and the phase $\delta$ of the modulation.  
Throughout this paper, the lattice constant is set equal to 1 and
we only consider the cases when $Q/2\pi$ takes rational numbers.
The above Hamiltonian may be written symbolically as
\begin{equation}
\label{eq:Hquad}
H=\sum_{i, j=1}^N\left[c_i^{\dagger}A_{ij}c_j+\frac{1}{2}\left(c_i^{\dagger}B_{ij}c_j^{\dagger} +c_iB_{ji}c_j   \right) \right],
\end{equation}
where $A$ and $B$ respectively are real symmetric and real skew-symmetric matrices
with $A_{i,i}=-2h\cos(Qi+\delta)$, $A_{i,i+2}=A_{i+2,i}=t$, $B_{i,i+2}=-B_{i+2,i}=t$, and 0 otherwise.
Note that a similar Hamiltonian with nearest-neighbor interaction
has been studied in Ref.~\citen{Lang-2012}.
The next-nearest-neighbor part of the Hamiltonian [Eq.~(\ref{eq:Hth})] 
(i.e., $H$ with $h=0$) is recast
by the Jordan--Wigner transformation\cite{Lieb-1961}
\begin{equation}
c_i=\prod_{j=1}^{i-1}(-\sigma_j^z)\,\sigma_i^-,\quad c_i^{\dagger}=\prod_{j=1}^{i-1}(-\sigma_j^z)\,\sigma_i^+
\end{equation}
into a spin model with the three-spin interaction (the cluster model)\cite{Suzuki-1971,Skrovseth-2009,Smacchia-2011}
\begin{equation}
\label{eq:clustermodel}
H_{\rm C}=-\sum_{i=1}^{N} t\sigma_i^x\sigma_{i+1}^z\sigma_{i+2}^x,
\end{equation}
where $\sigma_i^{\alpha}$ $(\alpha=x,y,z)$ are the Pauli matrices at site $i$.
For open boundary condition, we take $\sigma_{N+1}^\alpha=\sigma_{N+2}^\alpha=0$ ($\alpha = x,y,z$).
The three-spin interaction in Eq.~(\ref{eq:clustermodel}) and the ground state of the Hamiltonian $H_{\rm C}$
are respectively called the cluster interaction
(or the cluster stabilizer) and the cluster state in quantum-information science.\cite{Raussendorf-2001,Raussendorf-2003,Fujii-2013}
On the other hand, the modulating chemical potential corresponds, in the spin language, to an inhomogeneous magnetic field:
\begin{equation}
\label{eq:magneticfield}
H_{\rm field}=-\sum_{i=1}^{N} h\cos(Qi+\delta)\sigma_i^z.
\end{equation}
This field makes the spins polarize along the $z$-direction and locally changes the topological properties of the cluster phase.
The wave number $Q$ determines the distance between the nodes where the topological properties are scarcely affected.
The phase $\delta$ shifts the position of the nodes.

\begin{figure}
 \begin{center}
  \includegraphics[width=86mm]{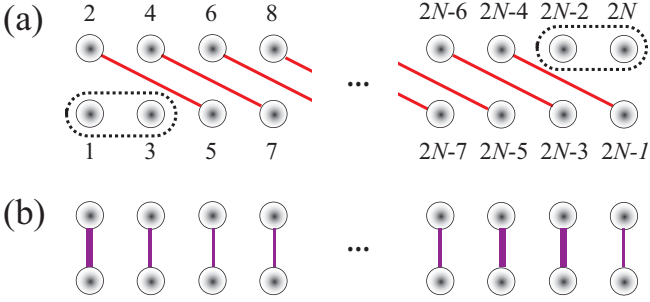}
 \end{center}
 \caption{(Color online)
 (a), (b) Schematic representation of the interactions in Eq.~(\ref{eq:Hquad}) by the pairing of the Majorana fermions.
 Fermion on each site is decomposed into two Majorana fermions (white circles).
 (a) depicts the next-nearest-neighbor interaction.
 The Majorana fermions enclosed by dotted circles are not contained in the Hamiltonian.
 Thus forming the gapless edge modes.
 (b) represents the modulating chemical potential (or the magnetic field).
 The strength is shown by the thickness of the bond.
 }
 \label{fig:Fig2}
\end{figure}

The physical properties of the ground state of the model defined by Eq.~(\ref{eq:Hquad}) become clear in the Majorana representation.\cite{Kitaev-2001}
The Majorana fermions \{$\bar{c}_i$\} constitute the real and imaginary parts of the spinless fermions \{$c_i$\}:
\begin{equation}
\bar{c}_{2i-1}=c_{i}^{\dagger}+c_i,\quad \bar{c}_{2i}=i\,(c_i-c_{i}^{\dagger}),\quad i=1, 2, \dots, N.
\end{equation}
The standard anticommutation relations of \{$c_i$\}, \{$c_i^{\dagger}$\} translate into
\begin{equation}
\bar{c}_{i}=\bar{c}_{i}^{\dagger},\quad \{\bar{c}_{i},\ \bar{c}_{j}\}=2\delta_{ij}.
\end{equation}
Let us introduce a vector $\bar{c}=(\bar{c}_{1},\bar{c}_{2},\dots,\bar{c}_{2N})^{\mathrm{T}}$ and a real skew-symmetric matrix $M$ with
$M_{2i-1,2i}=-M_{2i,2i-1}=h\cos(Qi+\delta), M_{2i,2i+3}=-M_{2i+3,2i}=t$, and 0 otherwise.
Then, in terms of the Majorana fermions, the model given by Eq.~(\ref{eq:Hquad}) is written compactly as
\begin{equation}
\label{eq:HGCmajorana}
H=\frac{i}{2}\bar{c}^{\mathrm{T}}M\bar{c}.
\end{equation}
For open boundary condition, we can easily find the number of the degenerate ground states
and the Majorana zero modes.\cite{Kitaev-2001,Ohta-2015,Fidkovski-2011}
When the hopping $t$ is dominant,
there are two unpaired Majorana fermions at each end of the system, as shown in Fig.~\ref{fig:Fig2} (a).
They form two zero-energy excitations localized at each
end of the system and consequently the four-fold degeneracy
in the ground states results.
On the other hand, when the amplitude $h$ of the modulation is dominant at each site,
the Majorana fermions ($\bar{c}_{2i-1}, \bar{c}_{2i}$) pair up locally;
in general, there are no unpaired Majorana fermions in the system [Fig.~\ref{fig:Fig2} (b)] and the ground state is unique.
Since time-reversal operation acts on the Majorana fermions as
\begin{equation}
\bar{c}_{2i-1} \to \bar{c}_{2i-1}, \quad \bar{c}_{2i} \to -\bar{c}_{2i},
\end{equation}
we can readily verify that the fermion Hamiltonian $H$ [Eq.~(\ref{eq:Hth}) or (\ref{eq:HGCmajorana})]
is time-reversal invariant supporting an integer number of Majorana edge modes.

\section{\label{sec:Methods}Methods}
We analyze the model [Eq.~(\ref{eq:Hquad})] by using the exact diagonalization method.\cite{Lieb-1961}
The model [Eq.~(\ref{eq:Hquad})] which is quadratic in fermion operators can be diagonalized as
\begin{equation}
\label{eq:HGCBogoliubov}
H=\sum_{\mu=1}^{N} \epsilon_{\mu}\left(\eta_{\mu}^{\dagger}\eta_{\mu}-\frac{1}{2}\right),\quad \epsilon_{\mu}\geq0
\end{equation}
by the Bogoliubov transformation
\begin{equation}
\eta_{\mu}=\sum_{i=1}^{N}\left[\frac{\phi_{i\mu}+\psi_{i\mu}}{2}c_i+\frac{\phi_{i\mu}-\psi_{i\mu}}{2}c_i^{\dagger}  \right].
\end{equation}
The matrices $\phi$, $\psi$ of order $N$ are the solution of the simultaneous equations:
\begin{subnumcases}
{}
\epsilon_{\mu}\psi_{i\mu}=\sum_{j=1}^{N}(A+B)_{ij}\phi_{j\mu}, \\
\epsilon_{\mu}\phi_{i\mu}=\sum_{j=1}^{N}(A-B)_{ij}\psi_{j\mu} \quad (i,\mu=1,2,\dots,N).
\end{subnumcases}
The eigenenergies \{$\epsilon_{\mu}$\} in Eq.~(\ref{eq:HGCBogoliubov})
are labelled in ascending order; $\epsilon_1 \le \epsilon_2 \le \cdots \le \epsilon_N$.
The column vectors of $\phi$, $\psi$ correspond to the real-space amplitudes of the Majorana fermions.
If all the eigenenergies $\epsilon_{\mu}$ are larger than 0,
the ground state is uniquely given by the Bogoliubov vacuum $\left| {\rm vac} \right\rangle$
satisfying $\eta_{\mu} \left| {\rm vac} \right\rangle = 0$ for all $\mu$.
If $\epsilon_1$ equals 0, there is a Majorana zero mode:
The state with the zero-energy Bogoliubov quasiparticle $\eta_{1}^{\dagger}\left| {\rm vac} \right\rangle$
is also a ground state of the model given by Eq.~(\ref{eq:Hquad}).
A similar argument applies to the case with more Majorana zero modes.\cite{Ohta-2015,Ohta-2016}

The topological phases are characterized by non-local correlation functions,
which, in our case, detect the pattern of the Majorana correlation.\cite{Ohta-2015,Ohta-2016}
Here we show how to calculate the following string correlation function $O_{\rm XZX}(L)$ 
of distance $L$:\cite{Smacchia-2011}
\begin{align}
\label{eq:stringcorrelationXZX}
O_{\rm XZX}(L)=&(-1)^L \left\langle\sigma_1^x\sigma_2^y\left(\prod_{i=3}^{L-2}\sigma_i^z\right)\sigma_{L-1}^y\sigma_L^x\right\rangle\\
\sim& \left\langle \prod_{j=1}^{L-2}\bar{c}_{2j}\bar{c}_{2j+3}\right\rangle.
\end{align}
In the ground state of the model $H_{\rm C}$ defined by Eq.~(\ref{eq:clustermodel}), 
$\langle \bar{c}_{2j}\bar{c}_{2j+3} \rangle$ is finite and so 
is the string order parameter $O_{\rm XZX}=\lim_{L \to \infty} O_{\rm XZX}(L)$.  
In the following, we take the expectation values with respect to the Bogoliubov vacuum 
$\left| {\rm vac} \right\rangle$.
A phase characterized by the non-vanishing string order parameter $O_{\rm XZX}$ is generally called the cluster phase.
Note that the ferromagnetic phase of the Ising Hamiltonian
\begin{equation}
H_{\rm Ising}=\sum_{i=1}^{N} \sigma_{i}^{x}\sigma_{i+1}^{x}
\end{equation}
also exhibits topological properties in the fermion representation;
the ground state has a pair of Majorana fermions at the ends of the system.\cite{Kitaev-2001}
The usual spin-spin correlation function which takes a finite value in the long-distance limit
is transformed, by the Jordan--Wigner transformation, into the following fermionic correlation function:
\begin{align}
\label{eq:stringcorrelationXX}
O_{\rm XX}(L)=&\left\langle\sigma_1^x\sigma_L^x\right\rangle\\
\sim& \left\langle \prod_{j=1}^{L-1}\bar{c}_{2j}\bar{c}_{2j+1}\right\rangle.
\end{align}
Since $O_{\rm XX}(L)$ has a non-local form in the fermion language,
we also call it the string correlation function in the following.
Table~\ref{table:BE} summarizes the relationship between the fermion and spin representations.

\begin{figure}
 \begin{center}
  \includegraphics[width=86mm]{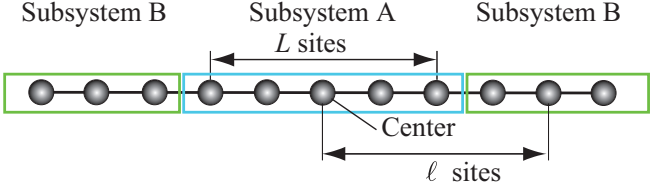}
 \end{center}
 \caption{(Color online)
 We measure the length $L$ symmetrically around the center of the system and the length $\ell$ from the center
 to calculate the physical quantities.
 We calculate the string correlation functions of length $L$ or $\ell$ and take $L$ adjacent sites as the subsystem to calculate the ES.
 }
 \label{fig:Fig3}
\end{figure}

We may also characterize the topological phase by the entanglement properties.
To quantify entanglement which is a measure of non-local quantum correlation between two systems,
we calculate the entanglement entropy (EE) and the ES.\cite{Li-2008,Latorre-2004}
To this end, we divide the entire system into a subsystems A
with length $L$ centered in the whole system and the rest B (Fig.~\ref{fig:Fig3}).
We first calculate the eigenvalues \{$\lambda_{\nu}$\} $(\nu=1,2,\dots, 2^{L})$ 
of the reduced density matrix $\rho_{\rm A}$ of the subsystem A 
obtained by tracing out the subsystem B:
\begin{equation}
\rho_{\rm A}(L)={\rm Tr}_{\rm B} \, \rho,
\end{equation}
where $\rho$ is the density matrix of the ground state of the entire system.
The EE is defined as the von Neumann entropy of $\rho_{\rm A}$:
\begin{equation}
S(L)=-{\rm Tr} \left[ \rho_{\rm A}(L) \ln \rho_{\rm A}(L) \right].
\end{equation}
On the other hand, the ES is
given by the logarithm of the eigenvalues of $\rho_{\rm A}(L)$\cite{Li-2008}
\begin{equation}
\xi_{\nu} = -\ln \lambda_{\nu}.
\end{equation}
Because the ES contains more information than the von Neumann EE $S(L)$, 
it has been used to study the topological phases.
It detects the degree of freedom appearing at the entanglement cut of the system.
\cite{Pollmann-2010,Cirac-2011,Lou-2011,Fagotti-2011,Tanaka-2012,Qi-2012,Nonne-2013,Cincio-2013,Tanimoto-2015}
In our case, the edge modes are the Majorana zero modes.\cite{Ohta-2015,Ohta-2016}

\begin{table}
 \caption{\label{table:BE}
 The relation between the fermion and spin representations.
 We show the number of the Majorana zero modes, degree of degeneracy in the lowest entanglement level, and the winding number $W$
 in each phase.
 The ES and the winding number are calculated
 in Sec.~\ref{sec:Quantum phase transitions} and Sec.~\ref{sec:Topological invariant}, respectively.
 }
  \begin{tabular}{lccc}
  & cluster &   ferromagnetic & paramagnetic \\
  \hline 
  spin & $\langle O_{\rm XZX}\rangle \neq 0$ & $\langle O_{\rm XX}\rangle \neq 0$ & disordered \\
  fermion & \begin{minipage}{20mm}
  topological \\
  (see Fig.~\ref{fig:Fig2} (a))
  \end{minipage} & topological & trivial  \\
  Majorana zero modes & 2 & 1 & 0 \\
  degeneracy in ES & four-fold & two-fold & not degenerate \\
  winding number $W$ & 2 & 1 & 0 
  \end{tabular}
\end{table}

\section{\label{sec:Quantum phase transitions}Quantum phase transitions}
We study the quantum phase transitions of the model defined by Eq.~(\ref{eq:Hquad})
when we vary (i) the amplitude $h$, (ii) the wave number $Q$, 
and (iii) the phase $\delta$ of the modulation $-2h\cos(Qi+\delta)$.
Depending on the phase $\delta$,
the nodes of the modulation exist on or between sites.
Near the nodes, the topological properties of the sites are only slightly affected.
The number of nodes depends on the wave number.
In fact, because of the cluster interaction in Eq.~(\ref{eq:clustermodel}),
the next-nearest-neighbor sites are interacting,
which is attributed to the wave length of the Majorana fermions (encoded in the real-space wave functions
$\phi$ and $\psi$).
Therefore it would be important to consider the two cases
when the nodes are separated by even or odd times the lattice constant.
In the following, we take $Q=\pi/2$ or $\pi/3$.
Note that the system with uniform field ($Q=\delta=0$) undergoes the quantum phase transition
to the quantum-paramagnetic phase\cite{Skrovseth-2009} at $h=1$.

\subsection{\label{sec:varyphase}Transitions with varying the phase $\delta$}
\begin{figure}
 \begin{center}
  \includegraphics[width=86mm]{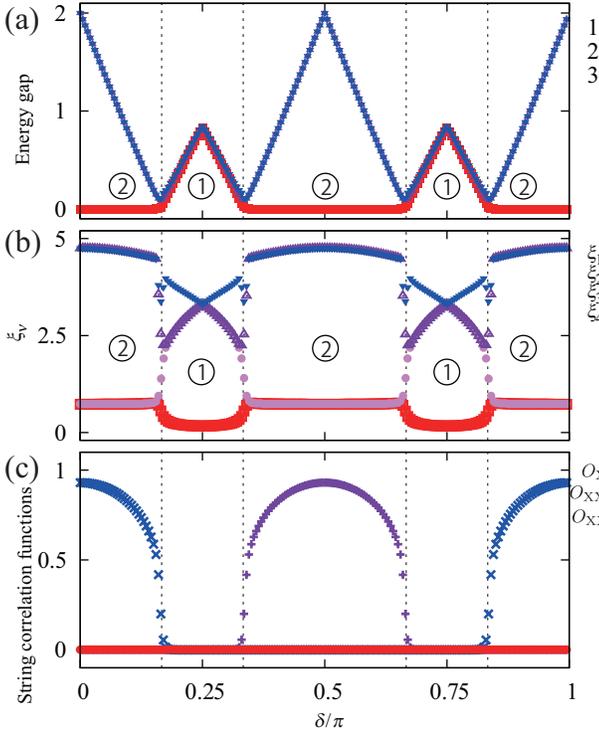}
 \end{center}
 \caption{(Color online)
 (a) The low-lying energy spectrum $E_{n}$ ($n=1,2,3$)
 with $N=201$ for $h=2$, $Q=\frac{\pi}{2}$.
 The energy gap closes at $\delta=\frac{\pi}{6},\frac{\pi}{3},\frac{2\pi}{3},\frac{5\pi}{6}$.
 (b) The ES with $N=201$ and $L=99$ for (c) $h=2$, $Q=\frac{\pi}{2}$.
 (c) The string correlation functions with $N=201$ for $h=2$, $Q=\frac{\pi}{2}$.
 The numbers in circles mean the degeneracy in the lowest energy/entanglement level.
 The degeneracies in the lowest level are 2 and 1 (no degeneracy)
 in the topological (ferromagnetic) and trivial (paramagnetic) phases, respectively.
 }
 \label{fig:Fig4}
\end{figure}

\begin{figure}
 \begin{center}
  \includegraphics[width=86mm]{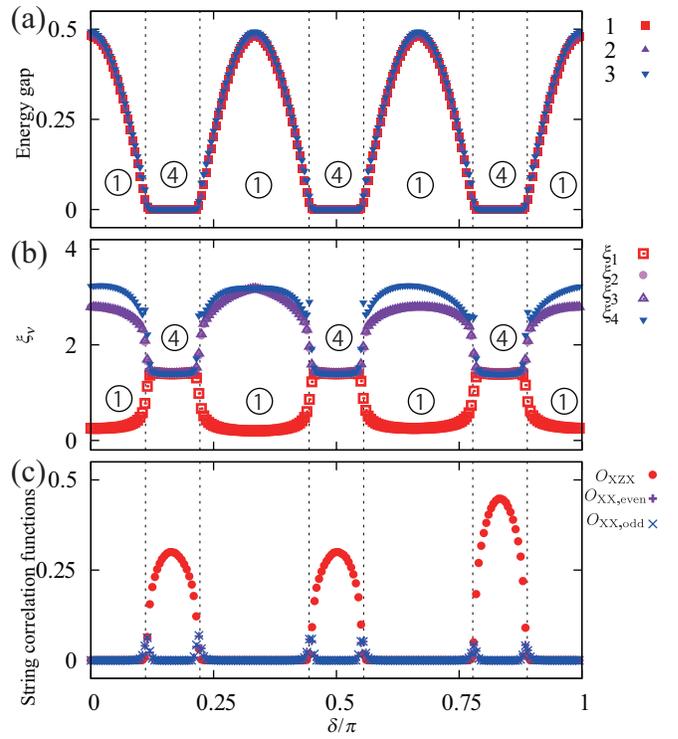}
 \end{center}
 \caption{(Color online)
 (a) The low-lying energy spectrum $E_{n}$ ($n=1,2,3$) with $N=201$ for $h=2$, $Q=\frac{\pi}{3}$.
 The energy gap closes at $\delta=\frac{\pi}{9},\frac{2\pi}{9},\frac{4\pi}{9},\frac{5\pi}{9},\frac{7\pi}{9},\frac{8\pi}{9}$.
 (b) The ES with $N=201$ and $L=99$ for $h=2$, $Q=\frac{\pi}{3}$.
 (c) The string correlation functions with $N=201$ for $h=2$, $Q=\frac{\pi}{3}$.
 The numbers in circles mean the degeneracy in the lowest level.
 The degeneracies in the lowest level are 4 and 1 (no degeneracy) in the cluster and paramagnetic phases respectively.
 }
 \label{fig:Fig5}
\end{figure}

We begin by studying the quantum phase transitions when we change the position 
of the nodes of the modulation by varying the phase $\delta$ of the modulation.
To characterize the phases, we calculate the energy gap and the ES,
which are plotted against the phase $\delta$ for $(h,Q)=(2,\pi/2)$ and $(2,\pi/3)$
in Figs.~\ref{fig:Fig4} (a, b) and \ref{fig:Fig5} (a, b), respectively.
We show the excitation energy between the vacuum and the three lowest-lying excited states 
$E_1=\epsilon_1$, $E_2=\epsilon_2$, $E_3=\min\{\epsilon_3,\epsilon_1+\epsilon_2\}$ in Figs.~\ref{fig:Fig4} (a) and \ref{fig:Fig5} (a),
where the dotted lines mark the critical points where the energy gap closes.
The numbers in the circles in Figs.~\ref{fig:Fig4} (a) and \ref{fig:Fig5} (a) [Figs.~\ref{fig:Fig4} (b) and \ref{fig:Fig5} (b)]
represent the degree of the degeneracy of the lowest energy level (entanglement level).
In each phase, the degree of degeneracy of the ground states coincides with that of the lowest level in the ES.
The Majorana fermions localized at the ends form non-local fermion excitations with zero energy,
which characterize the topological phases.
We thus confirm that the ES does reflect the fictitious edge modes at the entanglement cuts
even for the non-uniform system.
As the phase $\delta$ changes, we observe several quantum phase transitions.
At the critical points, the degeneracy structures disappear.
Because the degeneracy changes discontinuously across the critical points,
we can use the degree of degeneracy to characterize the phases and locate their boundaries.

To obtain more information on the nature of the phases, we calculate the string correlation functions.
In the calculation, the locations of the two end points (separated by a distance $L$) are crucial.
We take the two points in a symmetric way (i.e., $i=\frac{N-L+2}{2}$ and $i=\frac{N+L}{2}$) with respect to the center
as shown in Fig.~\ref{fig:Fig3}.
Now the system consists of $N=201$ sites obeying open boundary condition.
The $N$ sites of the system are labeled as in Fig.~\ref{fig:Fig1}.
In Figs.~\ref{fig:Fig4} (c) and \ref{fig:Fig5} (c),
we show the string correlation functions $O_{\rm XZX}(L)$ and $O_{\rm XX}(L)$ for $(h,Q)=(2,\pi/2)$ and $(2,\pi/3)$, respectively.
The $O_{\rm XZX}(L)$ in Eq.~(\ref{eq:stringcorrelationXZX}) is plotted for $L=99$.
We found that $O_{\rm XX}(L)$ strongly depends on the distance $L$
and we show both $O_{\rm XX}(L=99)$ (dubbed $O_{\rm XX,{\rm even}}$)
and $O_{\rm XX}(L=101)$ ($O_{\rm XX,{\rm odd}}$) in the same plots.
As we can see from Figs.~\ref{fig:Fig4} (c) and \ref{fig:Fig5} (c),
the string correlation functions have finite values in the phases where the ground states are degenerate:
$O_{\rm XZX} \neq 0, O_{\rm XX}=0$ ($O_{\rm XZX}=0, O_{\rm XX} \neq 0$) when the ground states are four-fold (two-fold) degenerate.
On the other hand, there is a phase where both $O_{\rm XZX}$ and $O_{\rm XX}$ vanish.
This phase is the paramagnetic phase because it is adiabatically connected to the disordered
(i.e., $\sigma^{z}$-ordered) phase of the transverse field Ising model.
Thus we may identify the phases whose degree of the ground-state degeneracy are 4, 2, and 1
with the cluster phase, the ferromagnetic phase, and the paramagnetic phase, respectively. 
In Sec.~\ref{sec:varyamplitude}, we will explain the appearance of the topological phases in Figs.~\ref{fig:Fig4} and \ref{fig:Fig5}.

\subsection{\label{sec:varyamplitude}Transitions with varying the amplitude $h$}
\begin{figure}
 \begin{center}
  \includegraphics[width=86mm]{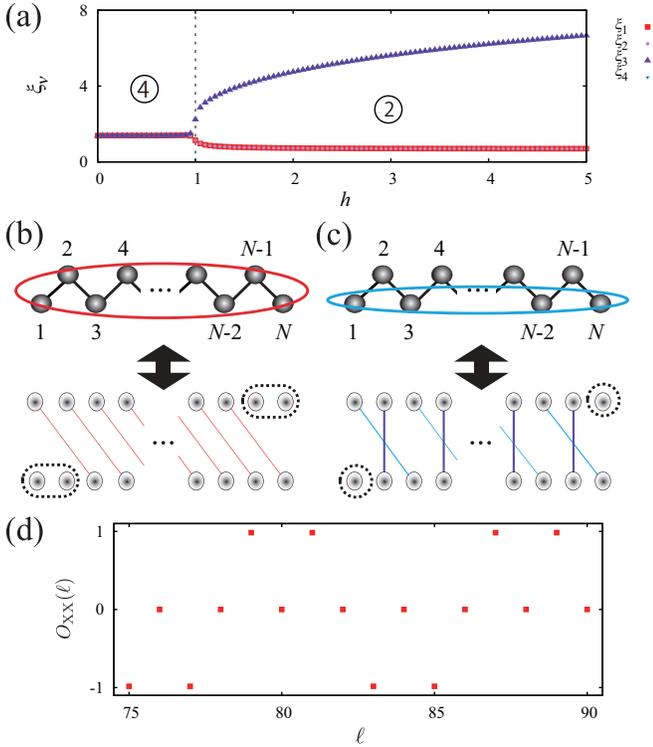}
 \end{center}
 \caption{(Color online)
 (a) The ES with $N=201$ and $L=99$ for $Q=\frac{\pi}{2}$ and $\delta=0$.
 There is a quantum phase transition between the cluster phase to the ferromagnetic phase at $h=1$.
 The numbers in circles represent the degeneracy in the lowest level.
 (b), (c) Schematic representation of the system.
 The Majorana fermions on the sites enclosed by the ellipses in the upper panels
 are paired with the next nearest ones as shown in the lower panels.
 (b) When the amplitude $h$ is smaller than 1,
 the entire system enclosed by red ellipse is still in cluster phase.
 (c) When the amplitude $h$ is larger than 1,
 the entire system is decomposed into paramagnetic part (even sites) and ferromagnetic part (odd sites) enclosed by blue ellipse
 coupled by the cluster interaction in Eq.~(\ref{eq:clustermodel}).
 (d) The length dependence of the string correlation function $O_{\rm XX}$ for $N=201$, $h=4$, $Q=\frac{\pi}{2}$, and $\delta=0$.
 The value is finite (0) when length $L$ is odd (even).
 }
 \label{fig:Fig6}
\end{figure}

In this subsection, we study the quantum phase transitions and the stability of the Majorana edge modes
by keeping track of the change in the ES when we vary the amplitude $h$ of the modulation.
For the wave number $Q=\frac{\pi}{2}$ and the phase $\delta=0$,
we plot the low-lying entanglement levels $\xi_1, \dots, \xi_4$
against the amplitude $h$ in Fig.~\ref{fig:Fig6} (a).
As the amplitude $h$ increases from 0 to 5, the lowest four-fold-degenerate level of the ES
splits at $h=1$ into a pair of two-fold-degenerate ones.
This corresponds to the topological phase transition from the cluster phase to the ferromagnetic phase.
We give the detailed explanation in the following.
With the wave number $Q=\frac{\pi}{2}$ and the phase $\delta=0$, the local magnetic field has 
the form $h \cos(\frac{\pi i}{2})$
and takes the value $\pm h$ (0) at even (odd) sites.
When the amplitude $h$ is smaller than 1, the perturbation does not close the energy gap and
only modifies the wave functions of the Majorana zero modes.
In fact, we can see in Fig.~\ref{fig:Fig6} (a) the four-fold degeneracy in the ES for $h<1$.
Thus the entire system is still in the topological cluster phase as is illustrated in Fig~\ref{fig:Fig6} (b).

On the other hand, this is not the case when the amplitude $h$ is larger than 1.
It is helpful to consider the case of strong field $h \gg 1$.
Then, the magnetic field on even sites is strong enough to destroy the ordered state,
while the odd sites hardly feel the magnetic field.
In this case, the lowest level of the ES are two-fold degenerate and the $O_{\rm XX,{\rm odd}}$ is finite, while $O_{\rm XX,{\rm even}}$ equals 0.
This indicates that the entire system is decomposed into
the trivial (paramagnetic) part (even sites) and the topological (ferromagnetic) part (odd sites).
The fermions in each part are coupled to each other by the cluster interaction [Eq.~(\ref{eq:clustermodel})] as is shown in Fig.~\ref{fig:Fig6} (c).
To confirm this, we calculated the length dependence of the string correlation function $O_{\rm XX}$
between ($\frac{N+1}{2}$) th site (the center) and ($\frac{N-1}{2}+\ell$) th site at $h=2$ [see Figs.~\ref{fig:Fig3} and \ref{fig:Fig6} (d)].
When $\ell$ is odd, $O_{\rm XX}$ has a finite value.
On the other hand, it is 0 when $\ell$ is even.
As the center site belongs to the odd sites,
this means that the even sites and odd sites are not correlated.
The spins on the even sites are fully polarized along the magnetic field whose direction depends on
the sign of the magnetic field.
By replacing $\sigma_{2i}^z$ with its expectation value $(-1)^{i}$,
the cluster interaction $\sigma_{2i-1}^x\sigma_{2i}^z\sigma_{2i+1}^x$ reduces to a next-nearest-neighbor interaction
$(-1)^{i}\sigma_{2i-1}^x\sigma_{2i+1}^x$ between the odd sites.
Because it alternates between a ferromagnetic interaction and antiferromagnetic one,
the spins on odd sites align as $\cdots \to \to \gets \gets \to \to \gets \gets \cdots$,
while those on the even sites are polarized in the $z$-direction.
We can see this in the oscillating structure in Fig.~\ref{fig:Fig6} (d).
Therefore we conclude that the entire system is decoupled into the spins on odd sites
that interact with each other via the alternating interactions
and the remaining paramagnetic part as shown in Fig.~\ref{fig:Fig6} (c).

\begin{figure}
 \begin{center}
  \includegraphics[width=86mm]{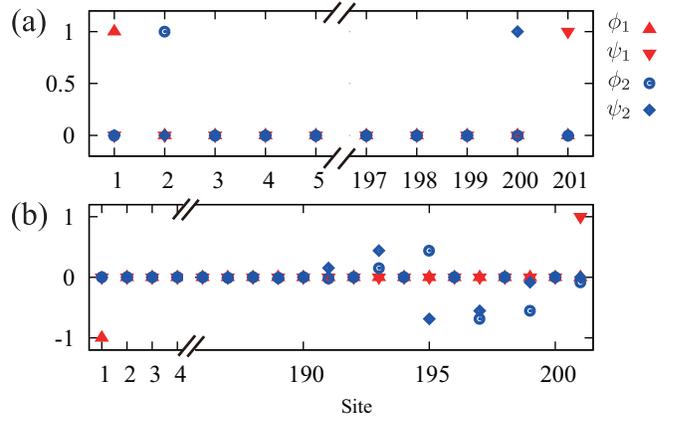}
 \end{center}
 \caption{(Color online)
 (a) The amplitudes of the Majorana fermions.
 $N=201$, $h=0$.
 (b) The amplitudes of the Majorana fermions.
 $N=201$, $h=3$, $Q=\frac{\pi}{2}$, and $\delta=0$.
 }
 \label{fig:Fig7}
\end{figure}

Next we turn our attention to the Majorana edge modes.
The amplitudes $\phi_{\nu}$ and $\psi_{\nu}$ of the $\nu$-th ($\nu=1,2$) Majorana fermion are shown in Figs.~\ref{fig:Fig7} (a) and (b).
The parameter in Fig.~\ref{fig:Fig7} (a) is $h=0$ for which the system belongs to the cluster phase.
The ground states are four-fold degenerate because of the two zero modes.
Note that each zero mode consists of a pair of Majorana fermions at the end of the system.
In the absence of the periodic modulation ($h=0$),
we confirm the existence of the localized Majorana fermions [Fig.~\ref{fig:Fig7} (a)].

On the other hand, the situation is quite different when $h=3$, $Q=\frac{\pi}{2}$, and $\delta=0$ as shown in Fig.~\ref{fig:Fig7} (b).
As in the previous case, the lowest two modes are localized, though the ground states are two-fold degenerate.
Because of the spatial periodicity of the modulation in our case,
the field at even sites is strong enough to make the Majorana fermions at even sites paired up locally,
while those at odd sites remain intact.
Therefore only one of the two zero modes is unaffected by the modulation [see Fig~\ref{fig:Fig6} (c)],
which is responsible for the two-fold degeneracy in the ES;
the other localized excitation acquires a finite energy.
This is why the ferromagnetic phase survives and the cluster phase does not appear in Fig.~\ref{fig:Fig4}.

\begin{figure}
 \begin{center}
  \includegraphics[width=86mm]{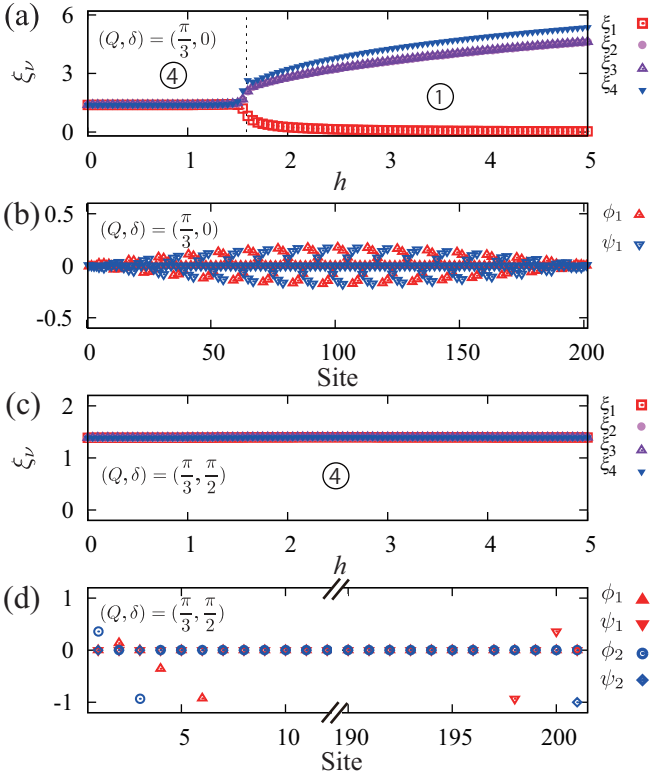}
 \end{center}
 \caption{(Color online)
 (a) The ES with $N=201$ and $L=99$ for $Q=\frac{\pi}{3}$ and $\delta=0$.
 The numbers in circles represent the degeneracy in the lowest level.
 There is a quantum phase transition between the cluster phase to the paramagnetic phase at $h = \sqrt[3]{4}$.
 (b) The amplitudes of the Majorana fermions.
 $N=201$, $h=3$, $Q=\frac{\pi}{3}$, and $\delta=0$.
 (c) The ES with $N=201$ and $L=99$ for $Q=\frac{\pi}{3}$ and $\delta=\frac{\pi}{2}$.
 The cluster phase is stable against the modulation.
 (d) The amplitudes of the Majorana fermions.
 $N=201$, $h=3$, $Q=\frac{\pi}{3}$, and $\delta=\frac{\pi}{2}$.
 }
 \label{fig:Fig8}
\end{figure}

Finally we study the stability of the cluster phase against the amplitude $h$ of the modulation.
In Fig.~\ref{fig:Fig8} (a), we show the plot of the ES vs the amplitude $h$ for the wave number $Q=\frac{\pi}{3}$ and the phase $\delta=0$.
The cluster phase remains stable when $h <\sqrt[3]{4}$.
Larger $h$ turns the ground state into the paramagnetic phase.
In Fig.~\ref{fig:Fig8} (b), the amplitudes of the Majorana fermions are shown
for the same ($Q, \delta$) and $h=3$ in the paramagnetic phase.
They now spread into the bulk.

For the wave number $Q=\frac{\pi}{3}$ and the phase $\delta=\frac{\pi}{2}$,
we show the plot of the ES vs the amplitude $h$ in Fig.~\ref{fig:Fig8} (c).
The degree of the degeneracy four in the ES does {\it not} depend on the amplitude $h$,
which suggests that the Majorana zero modes are stable even for large $h$.
This can be explained as follows.
There are sites which do not feel the magnetic field because of the spatial periodicity of the magnetic field.
Majorana zero modes can be localized at these zero points as shown in Fig.~\ref{fig:Fig8} (d).
This is why the Majorana zero modes are stable even for strong enough magnetic field.
A similar argument can be applied to the existence of the cluster phase in Fig.~\ref{fig:Fig5}.

\section{\label{sec:Topological invariant}Topological invariant}
In the previous section, we studied the topological phases in terms of the edge modes, the ES, and the non-local correlations
which are closely related to each other.
In this section, we try an alternative approach to study the topological property of the bulk system without boundary.
Specifically, we calculate a topological invariant to characterize the phases which we have identified above.
To calculate it, we impose periodic boundary condition on the system.
First, we perform the Fourier transformation. 
Since the $\mathbb{Z}_2$ index proposed in Ref.~\citen{Kitaev-2001}
cannot distinguish the phases with even (odd) numbers of edge modes,
we calculate the $\mathbb{Z}$ index.\cite{Wakatsuki-2014,Ciu-2015}
Rewriting the wave number $Q$ of the modulation as $2\pi \frac{p}{q}$ ($p,q$ are coprime integers),
we decompose the system into supercells with length $q$ and use the Fourier transformation with periodic modulation:
\begin{align}
c_{s,l}=\sqrt{\frac{q}{N}}\sum_{k=0}^{2\pi/q}c_{s,k}e^{ikql},
\end{align}
where $l \, (l=1,2, \dots, N/q)$ labels the supercells and $s \, (s=1,2,\dots,q)$ denotes a site in each supercell.\cite{Lang-2012}
The wave number $k$ is defined in the reduced Brillouin zone $[0, 2\pi/q]$.
Next, using these operators, we can construct new operators:
\begin{align}
\label{eq:gammaoperator}
\gamma_{2s-1}(k)=c_{s,k}+c_{s,-k}^{\dagger},\quad \gamma_{2s}(k)=-i \, (c_{s,k}-c_{s,-k}^{\dagger}).
\end{align}
For $k=0, 2\pi/q$, these operators satisfy the anticommutation relation of the Majorana fermions.
In this representation, we can rewrite our Hamiltonian in the following form:
\begin{align}
H=i\sum_{k=0}^{2\pi/q}\sum_{m,n=1}^{2q}B_{m,n}(k)\gamma_{m}(-k)\gamma_{n}(k),
\end{align}
\begin{equation}
  B = \left[
    \begin{array}{cc}
      0 & -v^{\dagger} \\
      v & 0
    \end{array}
  \right].
\end{equation}
This matrix $B$ is a $2q\times 2q$ complex skew-symmetric matrix and carries topological information.
The matrix elements of the submatrix $v$ are $v_{s,s}=-h \cos(\frac{s}{q}+\delta)$ for $s=1, 2, \dots, q$,
$v_{s,s+2}=-1$ for $s=1, 2, \dots, q-2$, and $v_{q-1,1}=v_{q,2}=-{\rm exp}(ikq)$, and 0 otherwise.
With the above setup in hand,
we are ready to calculate the topological invariant, called the winding number.\cite{Wakatsuki-2014,Ciu-2015}
We need to know how many times the eigenvalues \{$z_n(k)\} \, (n=1,\dots,q)$ of the submatrix $v$ go around the origin of the complex plane
as we change the wave number $k$ through the reduced Brillouin zone $[0, 2\pi/q]$:
\begin{align}
W=\sum_{n=1}^{q}\int_{0}^{2\pi/q}\frac{dk}{2\pi i}\partial_k {\rm ln} z_{n}(k).
\end{align}
We calculated it for the sets of parameters corresponding to the cluster, ferromagnetic, and paramagnetic phases
to find that $W$ is equal to 2, 1, and 0, respectively.
Therefore, we can see that the correspondence between the winding number of the bulk wave function
and the number of the Majorana edge modes holds even in the presence of the periodic modulation (Table~\ref{table:BE}).

\section{\label{sec:Summary}Summary}
In this paper, we have studied quantum phase transitions and the stability of the Majorana zero modes
of a generalized Kitaev model in one dimension where the chemical potential (or gate voltage) is spatially modulating.
First, we have studied quantum phase transitions taking place when we vary the phase of the modulation from the viewpoint of edge physics.
We have characterized each phase by the number of the Majorana zero modes, the ES, and the string correlation functions.
We have shown that the number of the Majorana zero modes is reflected in the degeneracy of the lowest level of the ES.
Second, focusing on the stability of the Majorana zero modes,
we have studied quantum phase transitions when we vary the amplitude of the modulation.
We have found a quantum phase transition between the topological phases and
shown that in certain cases the degeneracy of the ES does not change even when the amplitude of the modulation is sufficiently large.
Finally,
we have studied the topological properties of the bulk system
with an alternative approach assuming periodic boundary condition.
We have calculated the topological invariant,
which corresponds to the number of the zero-energy excitations that exist for open boundary condition.
We thus confirm the bulk-edge correspondence even in the presence of the spatially periodic modulation.

This work was supported by JSPS KAKENHI Grant Numbers 15K05211 (K.T.).
O.T. was supported by Grant-in-Aid for JPSJ Fellows.
The computations in the present work were carried out on the computers at Yukawa Institute for Theoretical Physics, Kyoto University.


\end{document}